\documentclass[aps,pra,twocolumn,showpacs,floatfix,superscriptaddress]{revtex4}
\usepackage{amsmath,hyperref}
\usepackage{amssymb}
\usepackage{graphics}
\usepackage{graphicx}
\usepackage{txfonts}
\usepackage{bbold}
\providecommand{\av}[1]{\left\langle #1\right\rangle}

\providecommand{\ket}[1]{|#1\rangle}
\providecommand{\bra}[1]{\langle#1|}
\providecommand{\kebra}[2]{\ket{#1}\bra{#2}}
\providecommand{\sprod}[2]{\langle#1|#2\rangle}
\providecommand{\tr}[1]{\textrm{tr}\{#1\}}
\providecommand{\tra}[1]{\textrm{tr}_a\{#1\}}

\providecommand{\ug}{\!=\!}
\providecommand{\meno}{\!-\!}
\providecommand{\bchi}{\boldsymbol{\chi}}

\providecommand{\eq}[1]{Eq.~(\ref{#1})}

\begin{document}
\title{Qubit-Controlled Displacements in Markovian Environments}
\author{Tommaso Tufarelli}
\affiliation{QOLS, Blackett Laboratory, Imperial College London, SW7 2BW, UK}
\affiliation{Department of Physics and Astronomy, University College London, Gower Street, London WC1E 6BT, UK}
\pacs{42.50.-p,03.65.Yz,03.67.-a}
\begin{abstract}
We study a particular form of interaction Hamiltonian between qubits and quantum harmonic oscillators, whose closed system dynamics results in qubit controlled displacement operations. We show how this interaction is realizable in many setups, including nanomechanical systems, ion traps, cavity QED and circuit QED, and in each context we provide quantitative estimates for the relevant parameters. The dynamics of the system is investigated through a master equation, including typical decoherence mechanisms resulting from the coupling of the qubit and oscillator to a thermal Markovian environment. We show how to solve the master equation by adopting a phase-space representation for the oscillator, and derive analytical and approximate solutions for many special cases of interest. Finally, our techniques are applied to a relevant example by studying the dynamics of qubit-oscillator entanglement and the preparation of oscillator states with negative Wigner function.
\end{abstract}
\maketitle
\section{Introduction}
Two objects that are ubiquitous in the Quantum Optics literature are the qubit (that is, a quantum two-level system) and the quantum harmonic oscillator (oscillator for brevity). 
From a mathematical point of view they are two of the simplest quantum systems, and yet they have been employed to model accurately a wide variety of physical phenomena. 
As a notable example, composite qubit-oscillator systems have been very successful tools for modelling and understanding the interaction between matter and quantized radiation \cite{loudon}.

Recently, however, substantial theoretical and experimental efforts have been directed in a somewhat opposite direction. Rather than using these two objects merely as theoretical tools to model ``naturally available'' systems, current research focuses on \textit{engineering} quantum systems such that their behaviour resembles that of qubits or oscillators as closely as possible. In addition, it is desirable to have accurate experimental control over these systems and their mutual interactions. This shift of perspective is partly due to the emergence of research subjects related to Quantum Technologies, such as Quantum Information \cite{nielsen}, and partly due to interest in the foundations of Quantum Mechanics. 

In this spirit, experimental platforms such as nanomechanical resonators \cite{nanomech}, ion traps \cite{ions-review}, cavity QED \cite{cavityQED-review} and circuit QED \cite{circuit-rev} have been capable of demonstrating a variety of qubit-oscillator models in the lab. Not only can these systems reproduce fundamental light-matter interactions on a larger scale and/or in completely different settings, they are also able to explore these interactions in parameter regimes that are unavailable at the microscopic level \cite{circuit-strong}, and even simulate qubit-oscillator Hamiltonians that have a qualitatively different form as compared to the ``natural'' interactions. This increased flexibility suggests the possibility of studying Hamiltonians that are alternative to e.g. the standard Jaynes-Cummings model of Quantum Optics \cite{j-c}, with a focus on their applicability to Quantum Technologies and to foundational studies of Quantum Mechanics.

Here we are concerned with one such model. We consider a particular form of interaction Hamiltonian between qubits and quantum harmonic oscillators, whose closed system dynamics generates \textit{qubit-controlled displacements}. That is, the time evolution resulting from our considered Hamiltonian shifts the average values of the position and momentum of the oscillator, by an amount that depends on the internal state of the qubit. This simple mechanism can establish strong quantum correlations between the two parts of the system \cite{tufa-4}, which in turn can be exploited in many applications that are of general interest for both technological and fundamental reasons. These include the preparation of non-classical oscillator states \cite{solano}, the preparation of entangled states that are robust against thermal noise \cite{Zheng,tufa-1}, the reconstruction of the quantum state of single oscillators \cite{tomo-prl,tufa-2} and oscillator networks \cite{tufa-3}, the realization of quantum communication tasks such as quantum teleportation or remote state preparation \cite{tufa-4}, the violation of Bell inequalities \cite{bell-padrenostro}, the thermometry of quantum oscillators \cite{padreparis}, the measurement of geometric phases \cite{padre2}, the probing of oscillator coherences \cite{armour} and many others. Even a proposal for full quantum computation has been put forward which relies on the same mechanism \cite{qbus}. In view of these and other proposals, the need has arisen to identify suitable experimental set-ups where the model can be implemented, and to investigate quantitatively how decoherence affects the realization of the desired tasks.

In this paper we give a contribution towards answering such questions. We show how to realize the proposed qubit-oscillator model in several realistic experimental platforms, namely nanomechanical systems, ion traps, cavity QED and circuit QED, and we solve the master equation resulting from the inclusion of common decoherence mechanisms affecting both qubits and oscillators. This can be of general usefulness to anyone interested in the study of the proposed system. We then provide two relevant examples in which our techniques are applied to study the influence of the thermal environment on physical quantities of interest. First we study the time evolution of qubit-oscillator entanglement, arguably the key resource in many of the applications described above. As a second example, we investigate the capability of the system to produce oscillator states with negative Wigner functions.

The paper is organized as follows. In Section~\ref{themodel} we introduce the model of interest, in particular presenting the general forms of the Hamiltonian and master equation that will be considered throughout the paper.  In Section~\ref{experiments} we provide a review of some experimental platforms where the model can be implemented, and for each considered system we include a detailed discussion of the necessary steps to obtain a master equation in the desired form. In Section~\ref{solving} we show how the master equation can be solved in its general form, and in particular we derive analytical expressions which are valid in many cases of interest. In Section~\ref{examples} our techniques are applied to study the dynamics of qubit-oscillator entanglement and the preparation of non-classical oscillator states, while in Section~\ref{fine} we draw our conclusions.
\section{The Model}\label{themodel}
We consider a qubit with ground state $\ket g$ and excited state $\ket e$, and an oscillator with annihilation operator $a$. The two interact via a Hamiltonian that conserves the excitations of the qubit. We suppose that such interaction is bilinear with respect to the qubit operator $\sigma_3$ [where ($\sigma_1,\sigma_2,\sigma_3$) is a right-handed tern of Pauli matrices] and the oscillator position, and we allow the coupling strength $g(t)$ to be a generic function of time. The Hamiltonian thus reads
\begin{equation}
H(t)\equiv g(t)\sigma_3\left(a {\rm e}^{-i\nu t}+a^\dagger {\rm e}^{i\nu t}\right),\label{CDH}
\end{equation}
where $\nu$ is a frequency that does not need to coincide with the natural frequency of the oscillator.
\subsection{Closed system dynamics}\label{closedsystem}
For the time being, let us assume that the system is closed, i.e., that its interaction with the external environment is negligible on the timescales of interest. Under this assumption, the density matrix obeys the Liouville equation \cite{messiah} 
\begin{equation}
	\dot\rho(t)=-i[H(t),\rho(t)].\label{liouville}
\end{equation}
This equation is formally solved by
\begin{equation}
	\rho(t)=U(t)\rho(0)U(t)^\dagger,
\end{equation}
where $\rho(0)$ is the initial state of the system, while $U(t)$ is the unitary time-evolution operator, which verifies the time-dependent Schr\"{o}dinger equation:
\begin{align}
	&\dot U(t)=-iH(t)U(t),\label{schrodinger}\\
	&U(0)=\mathbb1.
\end{align}
Solving Eq.~(\ref{schrodinger}) is in general difficult, since it involves the calculation of a time-ordered exponential. In our case, however, the time ordering operation just yields a global phase factor  which can be ignored for our purposes \cite{tufa-3}. The time evolution operator can thus be expressed as
\begin{equation}
	U(t)=\exp\left[-i\sigma_3\!\int_0^tdsg(s)\left(a{\rm e}^{-i\nu s}+a^\dagger {\rm e}^{i\nu s}\right)\right].
\end{equation}
Comparing the above expression with the definition of the displacement operator $D(\alpha)={\rm e}^{\alpha a^\dagger -\alpha^*a}$, we can recast the time evolutor in the compact form
\begin{align}
	&U(t)=D(\sigma_3\alpha(t)),\label{U_t}\\
	&\alpha(t)=-i\!\int_0^tdsg(s){\rm e}^{i\nu s}.\label{beta_t}
\end{align}
The operator of \eq{U_t} can be interpreted as a \textit{qubit-controlled displacement operator}, that is, a displacement operator for the oscillator, where the sign of the parameter $[\pm\alpha(t)]$ is determined by the eigenvalue of the qubit operator $\sigma_3$. 
\subsection{Open system dynamics}\label{open-sys}
So far we have neglected losses in the system, which allowed us to obtain a unitary time evolution. A realistic model, however, has to take into account the weak but inevitable coupling of the system to its surrounding environment. Here, we shall consider a widely applicable model of Markovian decoherence for both the qubit and the oscillator \cite{q-optics}. To take into account environmental effects, we write down a master equation by modifying \eq{liouville} as follows:
\begin{equation}
\dot\rho=-i[H(t),\rho]+L\rho+Q\rho.\label{master}
\end{equation}
The effect of decoherence is included through the following Lindblad terms, which are responsible for the non-unitarity of the dynamics:
\begin{align}
&L=\frac{\kappa}{2}(N_a+1)\mathcal D[a]+\frac{\kappa}{2}N_a\mathcal D[a^\dagger],\label{lind-1}\\
&Q=\frac{\Gamma_1}{2}(N_q+1)\mathcal D[\sigma^-]+\frac{\Gamma_1}{2}N_q\mathcal D[\sigma^+]+\frac{\Gamma_2}{4}\mathcal D[\sigma_3],\label{lind-3}
\end{align}
where $\mathcal D[\hat A]$, $\hat A$ being a generic operator, yields the Lindblad form
\begin{equation}
\mathcal D[\hat A]\rho=2\hat A\rho\hat A^\dagger-\hat A^\dagger\hat A\rho-\rho\hat A^\dagger\hat A. 
\end{equation}
In Eqs.~(\ref{lind-1}) and (\ref{lind-3}), the parameters $\kappa$, $\Gamma_1$ and $\Gamma_2$ quantify the effective coupling strength of the system to the environment. Typically, $\kappa$ and $\Gamma_1$ are associated to the exchange of excitations between system and environment, while $\Gamma_2$ to extra dephasing mechanisms affecting the qubit. Note that we have assumed that the environment acts independently on the two constituents of our system, and in particular, that the decoherence terms in the master equation \eqref{master} are not affected by the variations in the qubit-oscillator coupling. This treatment is consistent provided that the density of states of the environment is approximately constant in the frequency range relevant to the system [see e.g. Ref.~\cite{barrett} for a similar issue arising in driven solid state qubits]. The environment is assumed to be at thermal equilibrium at temperature $T$, hence $N_a$ and $N_q$ indicate the number of thermal excitations associated to the natural frequency of the oscillator and qubit respectively (dependence on the temperature $T$ has been left implicit). When the environment is modelled as a collection of harmonic oscillators, its number of thermal excitations at a given frequency is given by the Bose-Einstein distribution $N_j=({\rm e}^{\omega_j/T}-1)^{-1}$
where $\omega_j$, for $j=q,a$, is the natural frequency of either the qubit or the oscillator.

\section{Experimental Implementations}\label{experiments}
Before attacking \eq{master}, let us review a few experimental setups that can be used to implement our model. In the following examples, our aim will be to identify physical systems where the Hamiltonian (\ref{CDH}) and the master equation \eqref{master} can be reproduced with good approximation. Depending on the particular setup under examination, it can be the case that even though the ``bare" Hamiltonian is not in the desired form, it is still possible to obtain a Hamiltonian equivalent to Eq.~(\ref{CDH}) through appropriate unitary transformations and approximations. When this is the case, the applicability of the decoherence model of Eq.~(\ref{master}) has to be considered carefully, as one has to take into account also the effect that the mentioned transformations have on the decoherence terms of Eqs.~(\ref{lind-1}) and (\ref{lind-3}). We shall see that, for the cases considered in this paper, it is always possible to obtain a master equation in the same form as in Eq.~\eqref{master}, although the meaning of the various decoherence parameters might be different as compared to the standard scenario provided in Section~\ref{open-sys}.
\subsection{Flux qubit coupled to a nanomechanical resonator}\label{fei-xue-parameters}
It is known that our considered model can be implemented by coupling nanomechanical resonators to superconducting qubits \cite{rablzoller}. In particular, we shall focus here on the system put forward by Fei Xue {\em et. al.} \cite{flux-oscill}. In their theoretical proposal, the coupling between a superconducting flux qubit \cite{supcon1} and the fundamental flexural mode of a nanomechanical oscillator \cite{nanomech} is controlled by the amplitude of a transverse magnetic field, resulting in a Hamiltonian
\begin{align}
	&H_{\text{flux}}(t)=\frac{\varepsilon}{2}\sigma_3+\frac{\delta}{2}\sigma_1+\omega a^\dagger a+g(t)\sigma_3\left(a+a^\dagger\right),\label{Hsupercond}\\
	&g(t)=\eta B(t),\label{magnetic}
\end{align}
where $\varepsilon$ is the level splitting of the superconducting qubit, $\delta$ the tunnelling energy between the two levels, $\omega$ the mechanical oscillation frequency, $\eta$ a constant that depends on the specific system realization, and $B(t)$ the externally controlled magnetic field amplitude along an appropriate direction. A remarkable feature of this setup is that the three parameters $\varepsilon,\delta,g(t)$ can be tuned independently to each other, so that the Hamiltonian of Eq.~(\ref{CDH}) is easily obtained by operating in the regime $\delta\sim0$, and considering an interaction picture trasformation with respect to the free term $\tfrac{\varepsilon}{2}\sigma_3+\omega a^\dagger a$. As a typical set of parameters \cite{flux-oscill,supcon1,supcon-coherence,zoller-cooling}, we can take $\varepsilon\simeq2\pi\times10$GHz, $\omega\simeq2\pi\times100$MHz, $\eta\simeq2\pi\times0.8$MHz$/$mT. From the form of Eq. (\ref{magnetic}), we see that both the magnitude and sign of the coupling can be controlled over time by tuning the external magnetic field. Assuming a maximum magnetic field intensity $B_\textrm{max}\simeq10$mT, we get that the magnitude of the coupling strength $g(t)$ can vary between zero and $g_\textrm{max}\simeq2\pi\times8$MHz. For the decoherence parameters of the system we can take $\kappa\simeq0.3(\mu\text{s})^{-1}$, $\Gamma_1=\Gamma_2=2.5(\mu\text{s})^{-1}$, while the thermal excitations of the environment are given by $N_q\simeq0$, $N_a\simeq20$ at a temperature $T\sim100$mK. In units of the mechanical frequency, this parameter range yields a maximum coupling strength $g_\text{max}\simeq0.08\omega$, a qubit decay rate $\Gamma_1\simeq4\times10^{-3}\omega$ and dephasing rate $\gamma\simeq0.01\omega$, a mechanical coupling to the environment $\kappa\simeq5\times10^{-4}\omega$ and a heating rate $\kappa N_a\simeq0.01\omega$.
\begin{figure}
	\begin{center}
		\includegraphics[width=.65\linewidth]{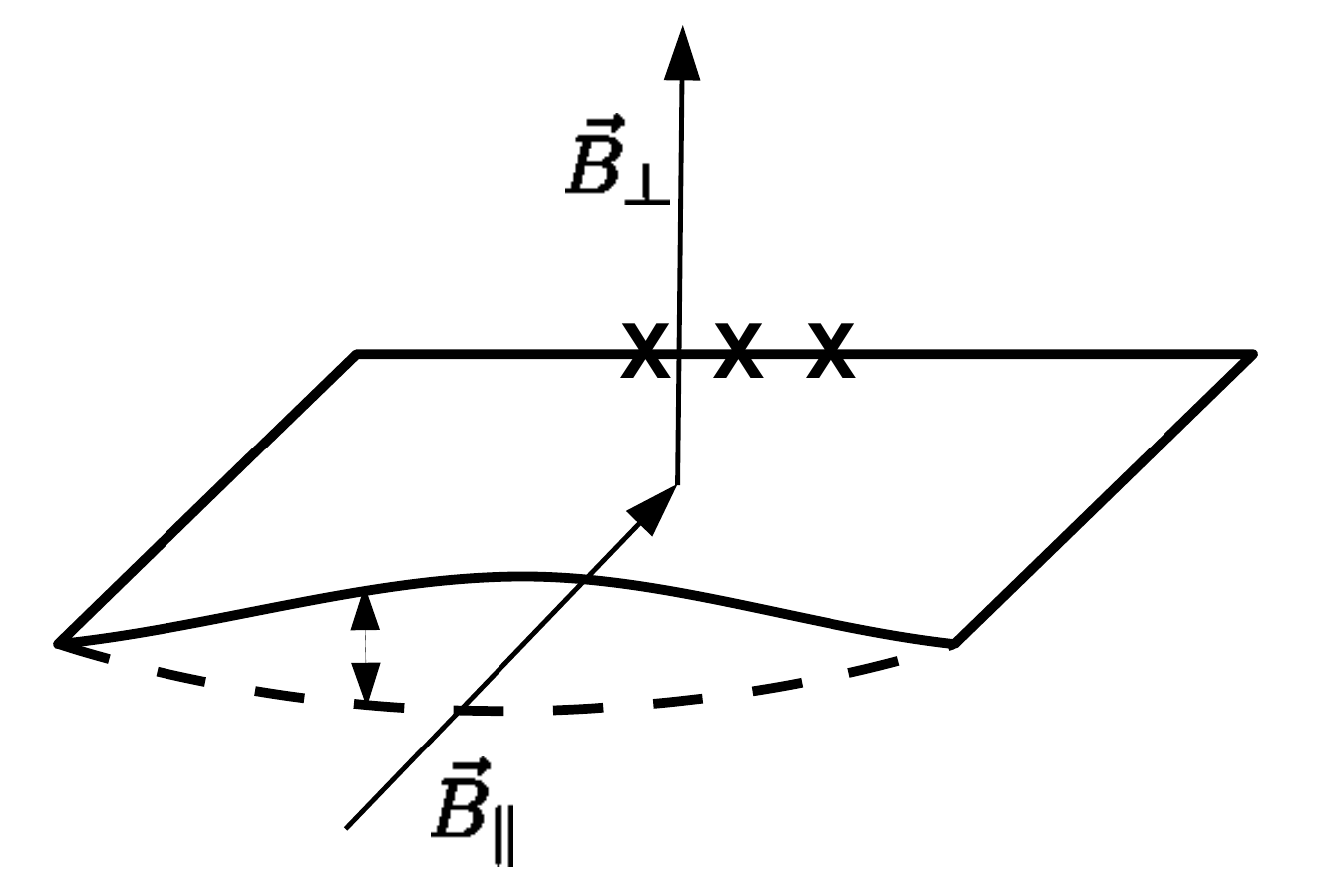}
	\end{center}
	\caption{Sketch of a flux qubit coupled to a nanomechanical resonator, as described in Ref.~\cite{flux-oscill}. In this picture, the flux qubit is given by a superconducting loop containing three Josephson Junctions (indicated with the symbol ``X'' in the picture), while the nanomechanical resonator is embedded in the loop, oscillating perpendicularly to the plane of the circuit. The magnetic field $\vec B_\parallel$, parallel to the plane of the circuit, and perpendicular to the direction of the mechanical oscillation, controls the strength of the qubit-oscillator interaction, while the perpendicular field $\vec B_\perp$ can be used to tune the free Hamiltonian of the qubit. See Ref.~\cite{flux-oscill} for more details.}
\end{figure}
\subsection{Trapped Ions}\label{ions-implementation}
Ion traps \cite{ions-review} constitute a rather mature experimental platform, which is especially suited for the implementation of our model. In this case two electronic levels of the ion provide the qubit, while the motion of the ion inside the trap can be modelled as a quantum harmonic oscillator with good approximation. For simplicity, we assume that we can consider only the motion of the ion along one of the trap axes \footnote{We shall assume that all the quantities of our interest are approximately constant within the range of motion of the ion in the $yz$ plane.}, say $x$. To couple the internal levels of the ion to its motion along the $x$ axis, we drive the system resonantly with a standing laser wave, in such a way that a node of the electric field coincides with the centre of the trap \cite{james-ions}. Correspondingly, the total Hamiltonian of the system reads
\begin{equation}
	H_{\text{ion}}(t)=\frac{\varepsilon}{2}\sigma_3+\omega a^\dagger a+2\Omega(t) \sigma_1\sin\left[kx_0\left(a+a^\dagger\right) \right]\cos{\varepsilon t},\label{ham-ions}
\end{equation}
where $\varepsilon$ is both the frequency of the ion electronic transition and that of the driving laser, $\omega$ the ion motional frequency along $x$, $\Omega(t)$ the Rabi frequency of the driving standing wave, with corresponding wavenumber $k$, $x_0$ the ground state spread of the ion along the $x$ axis. We allow the Rabi frequency to vary over time in a controlled way, provided that this variation is slow compared to the transition frequency\footnote{Since $\varepsilon$ is several orders of magnitude above the motional frequency $\omega$, this condition still allows to vary $\Omega(t)$ on a timescale that is fast compared to $\omega$.} $\varepsilon$. Taking an interaction picture with respect to $H_0=\frac{\varepsilon}{2}\sigma_3$, we get
\begin{align}
H'_{\text{ion}}(t)&\!=\!\omega a^\dagger a\!+\!\Omega(t) \left(\sigma_1\!+\!\sigma^+{\rm e}^{2i\varepsilon t}\!+\!\sigma^-{\rm e}^{\!-\!2i\varepsilon t}\right)\sin\left[kx_0\!\left(a\!+\!a^\dagger\right) \right].\label{ham-ions-intpic}
\end{align}
Note that $\varepsilon$ can be in the optical or infrared domain, while for the mechanical frequency a typical value is $\omega\simeq2\pi\times10$MHz. Therefore, if we are interested in the dynamics on a timescale of $\omega$, we can consider a \textit{Rotating Wave Approximation} (RWA), and neglect the fast oscillating terms at frequency $\pm2\varepsilon$ \cite{james-effective}. In addition, since typical ground state spreads for trapped ions are of the order of $x_0\sim1\div10$nm, while the wavenumber is of the order $k\sim10^7\text{m}^{-1}$ for optical transitions, it follows that $\eta_x=kx_0\ll1$. If in addition the motional state of the ion is in the Lamb-Dicke regime $\eta_x\sqrt{\av{(a+a^\dagger)^2}}\ll1$, we can approximate $\sin\left[\eta_x\left(a+a^\dagger\right) \right]\simeq\eta_x(a+a^\dagger)$. 
As a result, the Hamiltonian of Eq.~(\ref{ham-ions-intpic}) is well approximated by
\begin{align}
	&H'_{\text{ion}}(t)\simeq\omega a^\dagger a+g(t)\sigma_1\left(a+a^\dagger\right),\\
	&g(t)=\eta_x\Omega(t),
\end{align}
which can be transformed into \eq{CDH} via an interaction picture with respect to $\omega a^\dagger a$ and a qubit rotation $\sigma_1\rightarrow\sigma_3$. Note that there is a trade-off between the accuracy of the linear approximation of the sine, which improves as $\eta_x$ gets smaller, and the strength of the qubit-oscillator coupling, which is directly proportional to $\eta_x$. For example, considering a Lamb-Dicke parameter $\eta_x\simeq0.05$ can guarantee linearity of the coupling within a reasonable range of motion without weakening too much the coupling strength $g(t)$. By choosing a stable transition of the ion, decoherence of the qubit can be neglected, since the electronic levels involved can have lifetimes up to several seconds. Note however that the Rabi frequency scales as $\Omega\propto\sqrt{\mathcal I\varGamma}$, where $\mathcal I$ is the laser intensity and $\varGamma$ the decay rate of the excited state, so that stabler transitions require higher laser intensitites to be driven \cite{ions-review}. Rabi frequencies up to $\Omega_\text{max}\simeq2\pi\times1$MHz have been reported for dipole forbidden transitions with lifetime $\sim1$s \cite{rabi-freq}, which together with our choice $\eta_x\simeq0.05$ would allow $|g(t)|$ to vary between zero and a few percent of the motional frequency $\omega$. If the driving standing wave is realized by means of an optical cavity (operating in the semi-classical regime), this figure can be increased further. The time variability of the coupling can be preserved, provided that the cavity response time is small compared to the motional timescales of the ion. As anticipated, the availability of long internal lifetimes allows us to set $Q=0$ in the master equation (\ref{master}). On the other hand, motional decoherence is usually relevant, and typical heating rates in ion traps at room temperature are of the order $\kappa N_a\sim10^3\div10^4\text{s}^{-1}$. Thus, at a temperature $T\simeq300$K, we have $N_a\simeq6\cdot10^5$ for $\omega\simeq2\pi\times10$MHz, and we may take $\kappa\simeq0.02\text{s}^{-1}$. In units of the mechanical frequency, we have $\kappa\simeq6\times10^{-9}\omega$, $\kappa N_a\simeq2\times10^{-3}\omega$, and by considering a cavity-assisted driving we may assume $g_\text{max}\simeq0.05\omega$. Typically, the Lamb-Dicke regime in which our model is valid is reached by laser-cooling the ion to an excitation number much lower than $N_a$.
\begin{figure}
	\begin{center}
		\includegraphics[width=.48\linewidth]{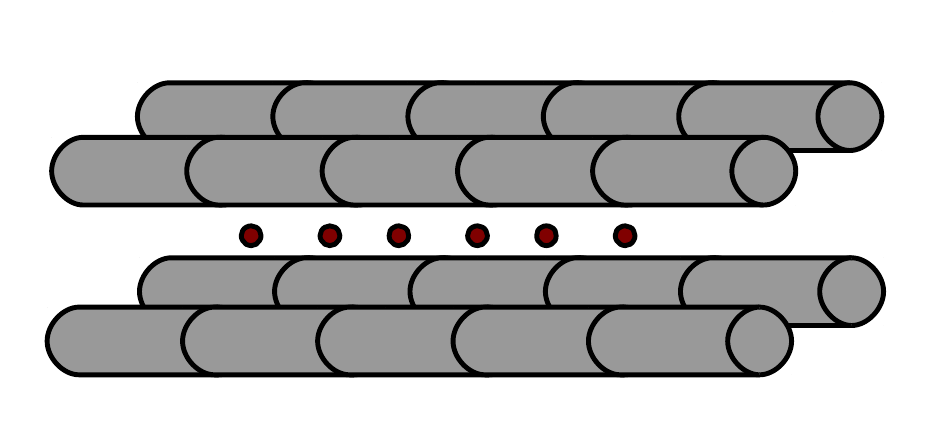}\hspace{.04\linewidth}\includegraphics[width=.45\linewidth]{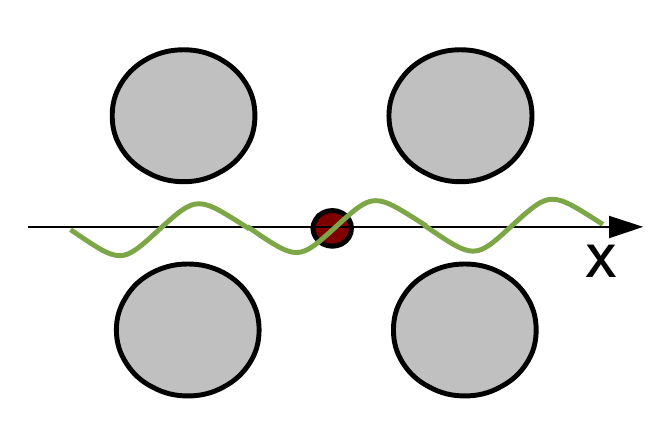}
	\end{center}
	\caption{Sketches of trapped ions. Left: several ions confined in a linear trap. Right: a laser standing wave is used to couple the motion of an ion along the $x$ axis with two of its internal levels. In order to realize the desired Hamiltonian the center of the ion motion has to coincide with a zero of the driving field \cite{james-ions}.}
\end{figure}
\subsection{Cavity QED}\label{cav-qed}
Cavity Quantum Electrodynamics (QED) has long been the paradigmatic quantum optical setup to investigate simple qubit-oscillator models \cite{cavityQED-review}. The qubit is provided by two electronic levels of an atom or ion (atom for brevity), while the oscillator represents a single confined mode of the electromagnetic field inside a high-finesse optical or microwave cavity (cavity mode for brevity). The atom can be either trapped inside the cavity, so that it interacts with the cavity mode continuously, or it can fly through the cavity and interact with its confined mode only for a limited time. The interaction between a two-level atom and a single-mode electromagnetic field is described by the Rabi model
\begin{equation}
	H_{\text{Rabi}}(t)=\frac{\varepsilon}{2}\sigma_3+\omega a^\dagger a+\bar g(t)\sigma_1\left(a +a^\dagger\right),\label{h-rabi}
\end{equation}
with $\varepsilon$ the level splitting of the atom, $\omega$ the frequency of the cavity mode, $\bar g(t)$ the interaction strength between atom and cavity mode at time $t$. The latter is usually well approximated by considering $\bar g(t)=g_0$ while the atom is within the effective volume of the cavity mode, and $\bar g(t)=0$ otherwise. Despite its rather simple form, the Rabi Hamiltonian of \eq{h-rabi} is in general difficult to treat analytically \cite{rabi-int}. However, if the atom and the cavity are resonant or near-resonant ($\varepsilon\simeq\omega$), and their coupling strength is weak compared to the uncoupled frequencies ($g_0\ll \varepsilon,\omega$), it is a good approximation to describe their interaction via the Jaynes-Cummings Hamiltonian \cite{j-c}:
\begin{equation}
	H_{JC}(t)=\frac{\varepsilon}{2}\sigma_3+\omega a^\dagger a+\bar g(t)\left(a^\dagger \sigma^-+a\sigma^+	\right).\label{h-jc}
\end{equation}
The Hamiltonian of Eq.~(\ref{h-jc}) has innumerable applications, nevertheless it leads to a very different dynamics from what we are pursuing in this paper. One possible way of obtaining a Hamiltonian of the form (\ref{CDH}) in cavity QED is to add a strong external driving to the atom \cite{solano}. Here, we adopt the complementary strategy of strongly driving the cavity, an approach that might be useful in situations where direct strong driving of the atom is not feasible. Consequently, we modify the Hamiltonian as
\begin{align}
	H_{\text{drive}}&=\frac{\varepsilon}{2}\sigma_3+\omega a^\dagger a+g_0\left(a\sigma^++a^\dagger\sigma^-\right)+\nonumber\\
	&+\Omega\left( a{\rm e}^{-i\varphi+i\varepsilon t}+a^\dagger {\rm e}^{i\varphi-i\varepsilon t}\right),\label{hdrive}
\end{align}
where $\Omega$ is the strength of the cavity driving, operated at resonance with the qubit frequency $\varepsilon$, $\varphi$ is the phase of the driving field at time $t=0$, and we considered the atom to be inside the cavity during the time of interest, so that $\bar g(t)=g_0$. 
Going to a rotating frame at the driving frequency $\varepsilon$, we get
\begin{equation}
	H_\text{drive}'=\delta a^\dagger a+g_0\left(a\sigma^++a^\dagger\sigma^-\right)+\Omega\left(a{\rm e}^{-i\varphi}+a^\dagger {\rm e}^{i\varphi}\right),
\end{equation}
where $\delta=\omega-\varepsilon$ is the detuning between atom and cavity. We see from this form of the Hamiltonian that the ``equilibrium position" of the oscillator in phase-space has been shifted. To find the new equilibrium position, it is necessary to take into account decoherence. The master equation reads
\begin{equation}
	\dot\rho=-i[H_\text{drive}',\rho]+L\rho+Q\rho,\label{master-QED}
\end{equation}
where $L$ and $Q$ keep the same form of Eqs.~(\ref{lind-1}) and (\ref{lind-3}) also in the rotating frame we are considering. We now multiply the master equation of Eq.~(\ref{master-QED}) by the displacement operator $D(\alpha_0)^\dagger$ on the left, and $D(\alpha_0)$ on the right, so that we realize the bosonic translation $a\rightarrow a+\alpha_0$. Imposing that the transformed master equation does not contain terms proportional to $(a\rho, a^\dagger \rho, \rho a, \rho a^\dagger)$, we get the value of $\alpha_0$ corresponding to the new phase-space centre of oscillation:
\begin{equation}
	\alpha_0=-\frac{\Omega {\rm e}^{i\varphi}}{\delta-i\frac{\kappa}{2}}.
\end{equation}
For simplicity, we choose the driving phase $\varphi$ such that $\alpha_0>0$. Correspondingly, the transformed master equation reads
\begin{equation}
	\dot\rho=-i[H''_\text{drive},\rho]+L\rho+Q\rho,
\end{equation}
where the new Hamiltonian is given by
\begin{equation}
	H''_\text{drive}= \delta a^\dagger a+g_0\left(a\sigma^++a^\dagger\sigma^-\right)+\Omega'\sigma_1\label{hdrive2}.
\end{equation}
Note that $\Omega'=g_0\alpha_0$ is an effective driving strength for the atom. We are now very close to getting our desired Hamiltonian. Taking an interaction picture with respect to the last term in \eq{hdrive2}, we obtain a modified master equation
\begin{align}
	\dot\rho &=-i[H_\text{drive}'''(t),\rho]+L\rho+Q'(t)\rho,\\
	H_\text{drive}'''(t)&=\delta a^\dagger a +\frac{1}{2}g_0\sigma_1\left(a+a^\dagger\right)+\frac{i}{2}g_0\tilde\sigma_2(t)\left(a-a^\dagger\right),\\
	Q'(t)&=\frac{\Gamma_1}{2}(N_q+1)\mathcal D\left[\tfrac{1}{2}(\sigma_1-i\tilde\sigma_2(t))\right]+\nonumber\\
	&+\frac{\Gamma_1}{2}N_q\mathcal D\left[\tfrac{1}{2}(\sigma_1+i\tilde\sigma_2(t))\right]+\frac{\Gamma_2}{4}\mathcal D[\tilde\sigma_3(t)],
\end{align}
where $\tilde\sigma_2(t)=\sigma_2\cos{2\Omega't}-\sigma_3\sin{2\Omega't}$ and $\tilde\sigma_3(t)=\sigma_3\cos{2\Omega't}+\sigma_2\sin{2\Omega't}$. If the strength of the effective driving is large enough, such that\footnote{Note that also the conditions $\Omega'\gg,\kappa,\Gamma_1,\Gamma_2$ are required to perform the RWA. These are automatically verified if we operate in the strong coupling regime $g_0\gg\kappa,\Gamma_1,\Gamma_2$.}\footnote{Assuming that $\delta\lesssim g_0$, the strong driving condition simply becomes the requirement that the equilibrium coherent state has a large amplitude: $|\alpha_0|\gg1$.} $\Omega'\gg g_0, \delta$, we can apply again the RWA, and neglect the fast-oscillating terms at frequency $\pm2\Omega'$ \cite{james-effective}. Applying this approximation to the Hamiltonian is straightforward, while some care has to be taken in applying it to the operator $Q'(t)$, as it contains terms proportional to $\cos^2{2\Omega't}$ or $\sin^2{2\Omega't}$, which are fast oscillating but do not average to zero. One can verify that the correct approximation is then
\begin{align}
	&\mathcal D\left[\tfrac{1}{2}(\sigma_1\pm i\tilde\sigma_2(t))\right]\simeq \frac{1}{4}\mathcal D\left[\sigma_1\right]+\frac{1}{8}\mathcal D\left[\sigma_2\right]+\frac{1}{8}\mathcal D\left[\sigma_3\right]\\
	&\mathcal D[\tilde\sigma_3(t)]\simeq\frac{1}{2}\mathcal D\left[\sigma_2\right]+\frac{1}{2}\mathcal D\left[\sigma_3\right],
\end{align}
resulting in
\begin{align}
	&Q'\simeq\frac{\gamma_1}{4}\mathcal D\left[\sigma_1\right]+\frac{\gamma_2}{4}\mathcal D\left[\sigma_2\right]+\frac{\gamma_2}{4}\mathcal D\left[\sigma_3\right],\\
	&\gamma_1=\Gamma_1\left(N_q+\tfrac{1}{2}\right), \qquad\gamma_2=\tfrac{\Gamma_1}{2}\left(N_q+\tfrac{1}{2}\right)+\Gamma_2.
\end{align}
One may also notice the relation 
\begin{equation}
	\frac{1}{2}(\mathcal D\left[\sigma_2\right]+\mathcal D\left[\sigma_3\right])=\mathcal D\left[\tfrac{\sigma_2+i\sigma_3}{2}\right]+\mathcal D\left[\tfrac{\sigma_2-i\sigma_3}{2}\right].
\end{equation}
Taking a further interaction picture with respect to $\delta a^\dagger a$, and a qubit rotation $\sigma_1\rightarrow\sigma_3$, we finally get the master equation:
\begin{align}
	&\dot\rho=-i[H_\text{cqed}(t),\rho]+L\rho+\bar{Q}\rho,\label{master-qed}\\
	&H_\text{cqed}(t)=g(t)\sigma_3\left(a {\rm e}^{-i\delta t}+a^\dagger {\rm e}^{i\delta t}\right),\label{h-cqed}\\
	&\bar{Q}=\frac{\gamma_2}{2}\left(\mathcal D[\sigma^+]+\mathcal D[\sigma^-]\right)+\frac{\gamma_1}{4}\mathcal D[\sigma_3].\label{q-cqed}
\end{align}
where $g(t)=g_0/2$ if the atom is inside the cavity, $g(t)=0$ otherwise. By comparing Eqs.\eqref{lind-3} and \eq{q-cqed}, we can see that in the latter case the heating and cooling rates of the qubit are equal, and that the role of decay and dephasing rates has been swapped. Another difference from the previous examples is that here the tunability of the coupling strength is restricted to switching between zero and a fixed value, essentially by controlling how much time the atom will spend inside the cavity. On the other hand, the atom-cavity detuning $\delta$, which in this case plays the role of the oscillator frequency, might be modified from experiment to experiment by applying a Stark shift to the atom\footnote{At the same time, the driving frequency has to be modified by the same amount, in order preserve resonance with the atom --- see \eq{hdrive}.}. By considering circular Rydberg atoms flying through superconducting microwave cavities \cite{cavity-entangled}, we can have coupling strengths $g=g_0/2\simeq 2\pi\times 25$kHz, together with low decoherence rates $\kappa\simeq 1(\text{ms})^{-1}$ for the cavity and $\Gamma_1\simeq0.03(\text{ms})^{-1}$ for the atoms \cite{cavityQED-review}. Considering cavity modes of frequency $\omega\simeq2\pi\times50$GHz operated at a temperature $T\simeq1.3\text{K}$ as reported in \cite{cavity-entangled}, we have $N_q\simeq0.15$ for the atoms, resulting in $\gamma_1=2\gamma_2\simeq0.04(\text{ms})^{-1}$. For the cavity mode the effective thermal occupation of the environment is slightly higher due to room temperature photons leaking inside the cryogenic chamber: $N_a\simeq0.8$. At resonance ($\delta=0$) the strong driving condition $\alpha_0\gg1$ implies that the cavity driving should be stronger than its decay rate: $\Omega\gg\kappa$. As here the coupling $g$ is fixed, it makes sense to re-express the other parameters in terms of this quantity: $\kappa\simeq6\times10^{-3}g$, $\kappa N_a\simeq8\times10^{-3}g$, $\gamma_1=2\gamma_2\simeq 2\times 10^{-4}g$, and a total dephasing rate $\gamma=\gamma_1+\gamma_2\simeq3\times 10^{-4}g$.
\begin{figure}[t]
	\begin{center}
		\includegraphics[width=.65\linewidth]{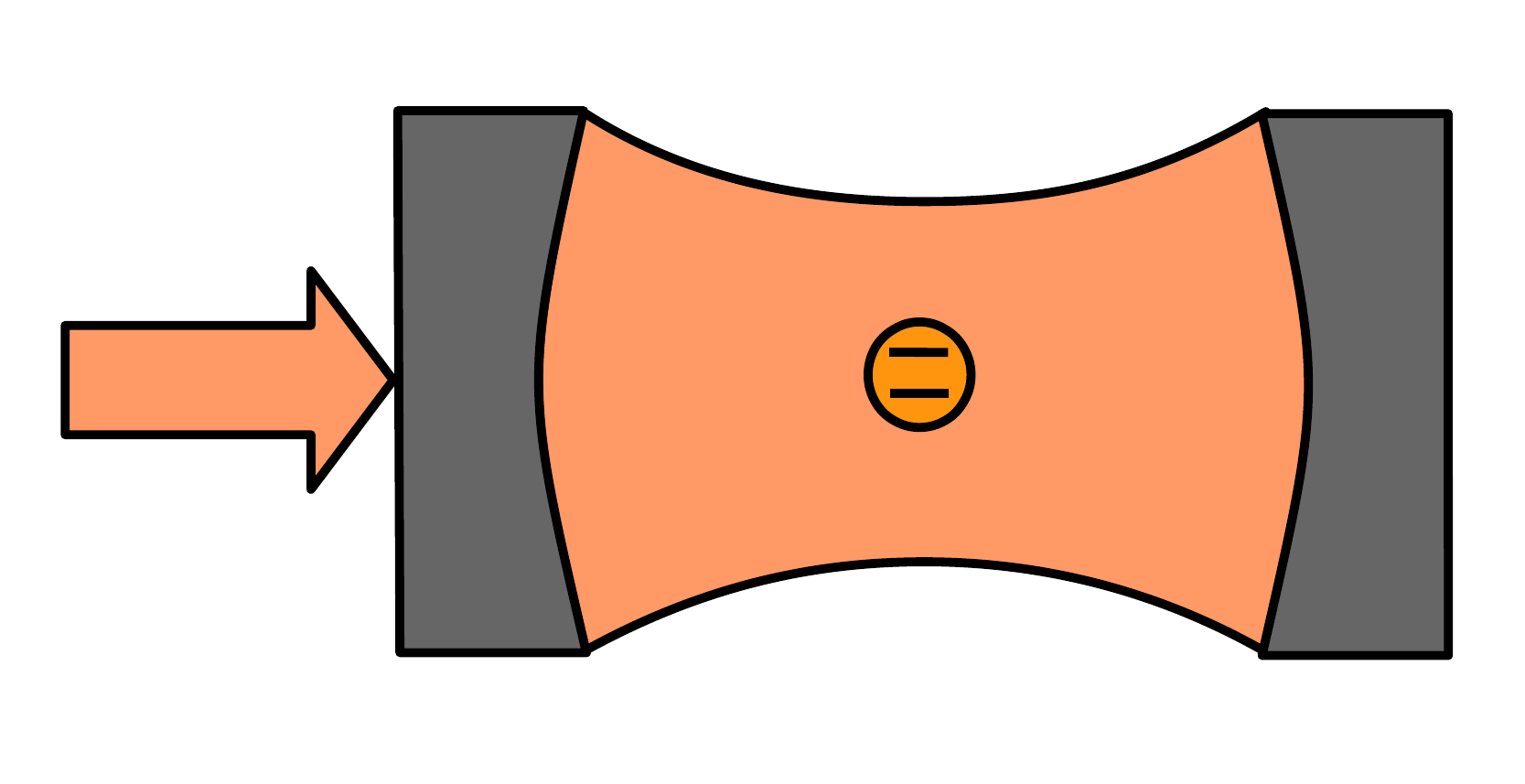}
	\end{center}
	\caption{Sketch of a cavity-QED setup. Two internal levels of an atom or ion are coupled to a mode of the electromagnetic field, confined between two highly reflective mirrors. The thick arrow on the left of the picture represents the external driving laser, necessary to obtain the desired Hamiltonian.}
\end{figure}
\subsection{Circuit QED}\label{exp-circuit}
In circuit QED \cite{circuit-rev}, the Rabi model of quantum optics is realized``on-chip", by coupling superconducting qubits (the artificial two-level atoms) to the quantized excitations of microwave resonators (i.e. microwave photons). In this architecture, light-matter interactions typical of cavity QED can be simulated in parameter ranges that are unavailable at the single-atom level \cite{circuit-strong}. Regimes where the Jaynes-Cummings model of \eq{h-jc} is valid are routinely achieved in circuit QED \cite{circuit-nature}, so that the same methods we discussed in Section~\ref{cav-qed} can be straightforwardly applied here to obtain the Hamiltonian of \eq{h-cqed}. Recently a novel superconducting qubit architecture has been proposed by the Houck group in Princeton, which allows for the independent tuning of both the qubit energy splitting and its coupling strength to the microwave photons \cite{houck-1}. Tunability of the coupling strength $g_0$ between a negligible value and $~2\pi\times45$MHz has been demonstrated in the Jaynes-Cummings regime, with the qubit and the cavity at resonance, $\omega=\varepsilon\simeq2\pi\times5.8$GHz \cite{houck-2}. The same group demonstrated that this tunability can be achieved on the timescale of few ns \cite{houck-3}. Consequently, this architecture should allow the realization of the Hamiltonian of \eq{h-cqed}, where $g(t)$ can be an experimentally controlled function of time. Typical decoherence parameters are $\kappa\simeq7.5(\mu\text{s})^{-1}$, $\Gamma_1\simeq0.6(\mu\text{s})^{-1}$, $\Gamma_2\simeq0.5(\mu\text{s})^{-1}$ while the system is operated at the low temperature $T\simeq25$mK, yielding $N_q=N_a\sim0$, and $\gamma_1\simeq0.3(\mu\text{s})^{-1}, \gamma_2\simeq0.65(\mu\text{s})^{-1}$ \cite{houck-2,houck-3}.

Even though what we have described until now applies to the Jaynes-Cummings regime, the versatility of the tunable qubit architecture should in principle allow to operate in a regime where the qubit energy splitting $\varepsilon$ is negligible as compared to the qubit-cavity coupling $\bar g(t)$. Then, the Rabi Hamiltonian alone would be sufficient to provide the required dynamics, without the need of an extra driving of the cavity. Indeed, by neglecting the qubit frequency in \eq{h-rabi}, we get
\begin{equation}
	H_\text{Rabi}(t)\simeq\omega a^\dagger a+\bar g(t)\sigma_1\left(a+a^\dagger\right),
\end{equation}
which is equivalent to \eq{CDH}, up to an interaction picture transformation.
\section{Solving the master equation}\label{solving}
The master equation \eqref{master} has been solved in the literature in several special cases of interest. These include a constant qubit-oscillator coupling \cite{armour,tufa-1} and a zero-temperature environment for the qubit \cite{tufa-2}. Here we discuss how to solve the master equation in its most general form. 
\subsection{C-Matrix representation}
To attack the master equation (\ref{master}), we adopt a phase-space representation for the oscillator degrees of freedom \cite{armour,tufa-1,tufa-2,tufa-3}. In particular, we shall simply extend the definition of the oscillator characteristic function \cite{q-optics} to a hybrid qubit-oscillator system. Note that the same idea can be applied to other phase-space representations, such as the Wigner function \cite{tufa-1,armour}. We define the C-Matrix as
\begin{equation}
\boldsymbol\chi(\beta,t)=\tra{\rho(t)D(\beta)},\label{cmatrix}
\end{equation}
where $\tra{\cdot}$ indicates partial tracing over the oscillator Hilbert space. Many properties of the characteristic function are easily extended to the C-Matrix. In particular, the knowledge of the C-Matrix is equivalent to the knowledge of the quantum state $\rho$, due to the inverse relationship $\rho =\tfrac{1}{\pi}\int d^2\beta \bchi(\beta)D(-\beta)$. The master equation \eqref{master} is equivalent to the following partial differential equation (PDE) for $\boldsymbol\chi$, where the complex parameters $\beta$ and $\beta^*$ are treated as independent variables \cite{q-optics}:
\begin{align}
&\dot{\boldsymbol\chi}=-ig(t)\left({\rm e}^{i\nu t}\partial_\beta-{\rm e}^{-i\nu t}\partial_{\beta^*}\right)[\sigma_3,\boldsymbol\chi]+\nonumber\\
&\phantom{\partial_t\chi=}+\!\frac{i}{2}g(t)\left({\rm e}^{-i\nu t}\beta\!+\! {\rm e}^{i\nu t}\beta^*\right)\{\sigma_3,\boldsymbol\chi\}\!+\!\mathcal L\boldsymbol\chi\!+\!Q\boldsymbol\chi.\label{master-chi}
\end{align}
In Eq.~\eqref{master-chi}, $Q$ has the same form as in \eq{lind-3}, $\{\cdot,\cdot\}$ is the anticommutator, while the operator responsible for the decoherence of the oscillator has now the differential form:
\begin{equation}
\mathcal L=-\frac{\kappa}{2}\left(\beta\partial_\beta+\beta^*\partial_{\beta^*}+2\Delta|\beta|^2\right),\label{lind-chi}
\end{equation}
with $\Delta=N_a+\tfrac{1}{2}$. Since Eq.~(\ref{master-chi}) only involves the qubit operators $\sigma_3,\sigma^+,\sigma^-$, it becomes convenient to decompose $\boldsymbol\chi$ in terms of the eigenstates of $\sigma_3$, as 
\begin{align}
\boldsymbol\chi(\beta,t)&=\chi_{ee}(\beta,t)\kebra{e}{e}+\chi_{gg}(\beta,t)\kebra{g}{g}+\nonumber\\
&+\chi_{eg}(\beta,t)\kebra{e}{g}\!+\!\chi_{ge}(\beta,t)\kebra{g}{e}.
\end{align}
As a result, we get the following system of PDEs for the C-Matrix elements:
\begin{align}
\dot\chi_{eg} &\ug 2ig(t)({\rm e}^{-i\nu t}\partial_{\beta^*}\meno {\rm e}^{i\nu t}\partial_\beta)\chi_{eg}\!+\!\mathcal L\chi_{eg}\!-\!\gamma\chi_{eg},\label{chi+}\\
\dot\chi_{ge} &\ug\!-\!2ig(t)({\rm e}^{-i\nu t}\partial_{\beta^*}\meno {\rm e}^{i\nu t}\partial_\beta)\chi_{ge}\!+\!\mathcal L\chi_{ge}\meno \gamma\chi_{ge}\label{chi-},\\
\dot\chi_{ee} &\ug ig(t)({\rm e}^{-i\nu t}\beta\!\!+\!{\rm e}^{i\nu t}\beta^*)\chi_{ee}\!+\!\mathcal L\chi_{ee}\!-\!\Gamma_c\chi_{ee}\!+\!\Gamma_h\chi_{gg}\label{chig},\\
\dot\chi_{gg} &\ug\!-\!ig(t)({\rm e}^{-i\nu t}\beta\!+\!{\rm e}^{i\nu t}\beta^*)\chi_{gg}\!+\!\mathcal L\chi_{gg}\!-\!\Gamma_h\chi_{gg}\!+\!\Gamma_c\chi_{ee}.\label{chie}
\end{align}
In the above equations, $\gamma=\Gamma_1(N_q\!+\!1/2)\!+\!\Gamma_2$ can be interpreted as the total dephasing rate of the qubit, while $\Gamma_c=\Gamma_1(N_q\!+\!1), \Gamma_h=\Gamma_1N_q$ are respectively its cooling and heating rates. Note that for the special cases of cavity-QED and circuit-QED, the decoherence parameters in Eqs.~\eqref{chi+}-\eqref{chie} are instead given by $\gamma\ug\gamma_1\!+\!\gamma_2\!=\!\tfrac{3}{2}\Gamma_1\left(N_q\!+\!\tfrac{1}{2}\right)\!+\!\Gamma_2$, $\Gamma_c\!=\!\Gamma_h\!=\!\gamma_2\!=\!\tfrac{\Gamma_1}{2}\left(N_q\!+\!\tfrac{1}{2}\right)\!+\!\Gamma_2$. 
\subsection{Off-diagonal terms}
We can see that Eqs.~(\ref{chi+}) and (\ref{chi-}) are in decoupled form, and in fact they can be solved analytically for any choice of the parameters. The corresponding solutions are given by
\begin{align}
	\chi_{eg}(\beta,t)&\ug\chi_{eg}(\beta {\rm e}^{\!-\!\frac{\kappa}{2}t}\meno\xi(t),0){\rm e}^{-\Delta\left(1\meno {\rm e}^{-\kappa t}\right)|\beta\meno\mu(t)|^2-\tau(t)}\label{chi+ansatz},\\
	\chi_{ge}(\beta,t)&\ug\chi_{ge}(\beta {\rm e}^{\!-\!\frac{\kappa}{2}t}\!+\!\xi(t),0){\rm e}^{-\Delta\left(1-{\rm e}^{-\kappa t}\right)|\beta+\mu(t)|^2-\tau(t)},\label{chi-ansatz}	
\end{align}
where
\begin{align}
	\xi(t)&\ug2i\int_0^t\mathrm{d}sg(s){\rm e}^{i\nu s-\frac{\kappa}{2}s}\label{xi_t},\\
	\mu(t)&\ug\frac{2i}{\sinh{\frac{\kappa}{2}t}}\int_0^t\mathrm{d}sg(s){\rm e}^{i\nu s}\sinh{\frac{\kappa}{2}s}\label{mu_t},\\
	\tau(t)&\ug\gamma t+\kappa\Delta\int_0^t\mathrm{d}s|\mu(s)|^2\label{nu_t}.
\end{align}
Some important applications such as the reconstruction of the oscillator state \cite{tufa-1,tufa-2} only make use of the two C-Matrix elements $\chi_{eg},\chi_{ge}$, so that in this case the above solutions allow to analyse exhaustively any parameter regime where our model is valid. On the other hand, it is obvious that the matrix elements $\chi_{eg},\chi_{ge}$ alone are not enough to describe the complete dynamics of the system, and in general the remaining elements $\chi_{ee},\chi_{gg}$ will be needed.
\subsection{In-diagonal terms}\label{complicanze}
Eqs.~(\ref{chig}) and (\ref{chie}) represent a system of two PDEs coupled to each other via the heating and cooling rates of the qubit, and their solution presents more difficulties. We shall show here how it is possible to reduce them to a system of ordinary differential equations (ODEs) in $t$, for each value of the complex parameter $\beta$. Analytical solutions are presented for the special case in which the heating rate vanishes.
\subsubsection{General treatment}
We shall start to build our solutions for the C-Matrix elements $\chi_{ee},\chi_{gg}$ by considering the uncoupled problem obtained by setting $\Gamma_h=0$ in Eq.~\eqref{chig} and $\Gamma_c=0$ in Eq.~\eqref{chie}. It can be checked directly that this is solved by
\begin{align}
\bar\chi_{gg}(\beta,t)&\ug {\rm e}^{\!-\!\Gamma_h t\!-\!\Delta\left(1\!-\!{\rm e}^{\!-\!\kappa t}\right)|\beta|^2-\lambda(t)\beta^*+\lambda(t)^*\beta}\chi_{gg}(\beta {\rm e}^{\!-\!\frac{\kappa}{2}t},0),\label{chig-homo}\\
\bar\chi_{ee}(\beta,t)&\ug {\rm e}^{\!-\!\Gamma_c t\!-\!\Delta\left(1\!-\!{\rm e}^{\!-\!\kappa t}\right)|\beta|^2+\lambda(t)\beta^*-\lambda(t)^*\beta}\chi_{ee}(\beta {\rm e}^{\!-\!\frac{\kappa}{2}t},0),\label{chie-homo}\\
\lambda(t)&\ug i\int_0^t\mathrm{d}sg(s){\rm e}^{i\nu s-\frac{\kappa}{2}(t-s)}.\label{lambda-sol}
\end{align}
Then, we can use the uncoupled solutions to write down an ansatz for the coupled problem, in the form
\begin{align}
	\chi_{gg}(\beta,t)&=\bar\chi_{gg}(\beta,t)\Phi_{gg}(\beta{\rm e}^{\!-\!\frac{\kappa}{2}t},t),\label{full-g}\\
	\chi_{ee}(\beta,t)&=\bar\chi_{ee}(\beta,t)\Phi_{ee}(\beta{\rm e}^{\!-\!\frac{\kappa}{2}t},t).\label{full-e}
\end{align}
After some manipulations it can be verified that the above expressions represent the exact solution to Eqs.~\eqref{chig} and \eqref{chie}, provided that, for each fixed value of $\beta$, $\Phi_{gg}(\beta,t)$ and $\Phi_{ee}(\beta,t)$ as functions of time solve the following system of ODEs
\begin{align}
\!\!\dot\Phi_{gg}(\beta,\!t)\!&=\!\Gamma_c\frac{\chi_{ee}(\beta,\!0)}{\chi_{gg}(\beta,\!0)}{\rm e}^{(\Gamma_h\!-\!\Gamma_c)t+2{\rm e}^{\frac{\kappa}{2}t}[\lambda(t)\beta^*\!-\!\lambda(t)^*\beta]}\Phi_{ee}(\beta,\!t),\label{ode-g}\\
\!\!\dot\Phi_{ee}(\beta,\!t)\!&=\!\Gamma_h\frac{\chi_{gg}(\beta,\!0)}{\chi_{ee}(\beta,\!0)}{\rm e}^{(\Gamma_c\!-\!\Gamma_h)t+2{\rm e}^{\frac{\kappa}{2}t}[\lambda(t)^*\beta\!-\!\lambda(t)\beta^*]}\Phi_{gg}(\beta,\!t),\label{ode-e}
\end{align}
with initial conditions $\Phi_{gg}(\beta,\!0)\!=\!\Phi_{ee}(\beta,\!0)\!=\!1$. In general, the above system of ODEs can be solved numerically for each value of $\beta$, which provides a net computational advantage as compared to attacking directly the system of PDEs \eqref{chig} and \eqref{chie}. We stress that, once the funcions $\Phi_{e,g}(\beta,t)$ are available, the substitution $\beta\to\beta {\rm e}^{-\frac{\kappa}{2}t}$ has to be performed before these are used to build the full solutions to Eqs.~\eqref{full-g} and \eqref{full-e}.
\subsubsection{Perturbative solution}
In many cases one is interested in interaction times such that $\Gamma_h t,\Gamma_c t\ll1$. Then, a reasonable approximation to the exact solution to Eqs.~\eqref{ode-g} and \eqref{ode-e} can be given at first order in time-dependent perturbation theory:
\begin{align}
	\Phi_{gg}(\beta,\!t)&\simeq1+\Gamma_c\frac{\chi_{ee}(\beta,\!0)}{\chi_{gg}(\beta,\!0)}\int_0^t{\rm d}s\,{\rm e}^{2{\rm e}^{\frac{\kappa}{2}s}[\lambda(s)\beta^*\!-\!\lambda(s)^*\beta]},\\
	\Phi_{ee}(\beta,\!t)&\simeq1+\Gamma_h\frac{\chi_{gg}(\beta,\!0)}{\chi_{ee}(\beta,\!0)}\int_0^t{\rm d}s\,{\rm e}^{2{\rm e}^{\frac{\kappa}{2}s}[\lambda(s)^*\beta\!-\!\lambda(s)\beta^*]}.
\end{align}
\subsubsection{Vanishing heating rate}
In the special case $\Gamma_h\simeq0$ the equations decouple, and in particular Eq.~\eqref{ode-e} simply gives $\dot\Phi_{ee}=0$, hence $\Phi_{ee}(\beta,t)=1$. Then, Eq.~\eqref{ode-g} is straightforwardly integrated as
\begin{equation}
	\Phi_{gg}(\beta,\!t)=1+\Gamma_c\frac{\chi_{ee}(\beta,\!0)}{\chi_{gg}(\beta,\!0)}\int_0^t{\rm d}s\,{\rm e}^{\!-\!\Gamma_cs+2{\rm e}^{\frac{\kappa}{2}s}[\lambda(s)\beta^*\!-\!\lambda(s)^*\beta]}.
\end{equation}
In applications where $\Gamma_h$ is interpreted as a physical heating rate [such as Sections~\ref{fei-xue-parameters} and \ref{ions-implementation}], this corresponds to considering an environment for the qubit that does not possess thermal excitations: $N_q\simeq0$. Note that while assuming this, we will still allow the possibility of having a thermally excited environment for the oscillator: $N_a\neq0$. This can be the case when the frequency of the qubit is several orders of magnitude higher than that of the oscillator (this is the case e.g. in flux qubits coupled to nanomechanical oscillators or ion traps, as discussed above). 
\section{Applications}\label{examples}
Having shown how to derive our model in realistic experimental setups and discussed how to solve its dynamics in the most general form, we move on to apply our techniques to two practical examples that can be of general interest: the study of entanglement in mixed qubit-oscillator states and the preparation of states with non-classical Wigner functions. To keep the mathematical complexity to a bare minimum, we shall consider a simple instance of our model: we take the Hamiltonian of Eq.~\eqref{CDH} with a constant coupling profile $g(t)=g_0$ and a vanishing oscillator frequency $\nu=0$. This can be the case, e.g., in cavity QED and circuit QED setups (Sections~\ref{cav-qed} and \ref{exp-circuit}).
In order to avoid the technicalities associated with the qubit heating and cooling rates (see Section~\ref{complicanze}) we set $\Gamma_h=\Gamma_c\simeq0$, and consider a qubit decoherence of pure dephasing: $\gamma\neq0$ (in fact, when considering the data presented in Sections~\ref{cav-qed} and \ref{exp-circuit} one might neglect qubit decoherence altogether, as a first approximation).
\subsection{Dynamics of qubit-oscillator entanglement}
\begin{figure}
	\begin{center}
		\includegraphics[width=.48\linewidth]{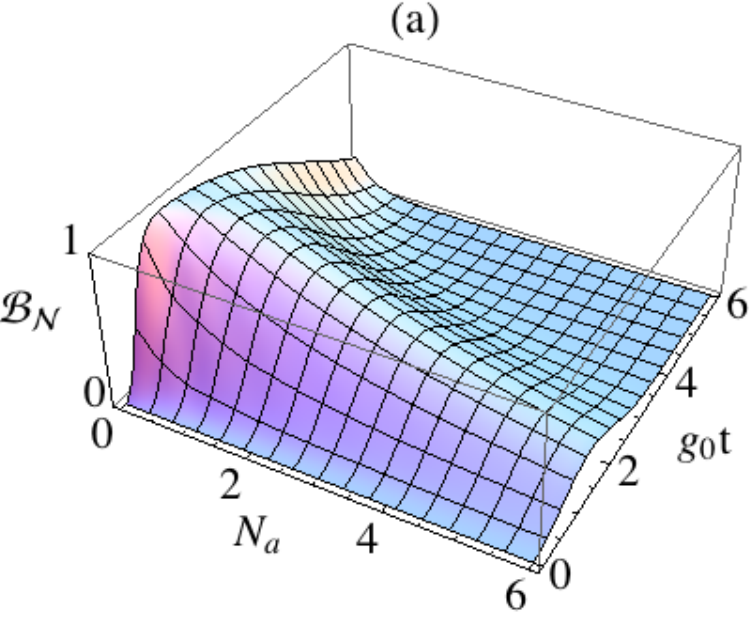}\hspace{.02\linewidth}		\includegraphics[width=.48\linewidth]{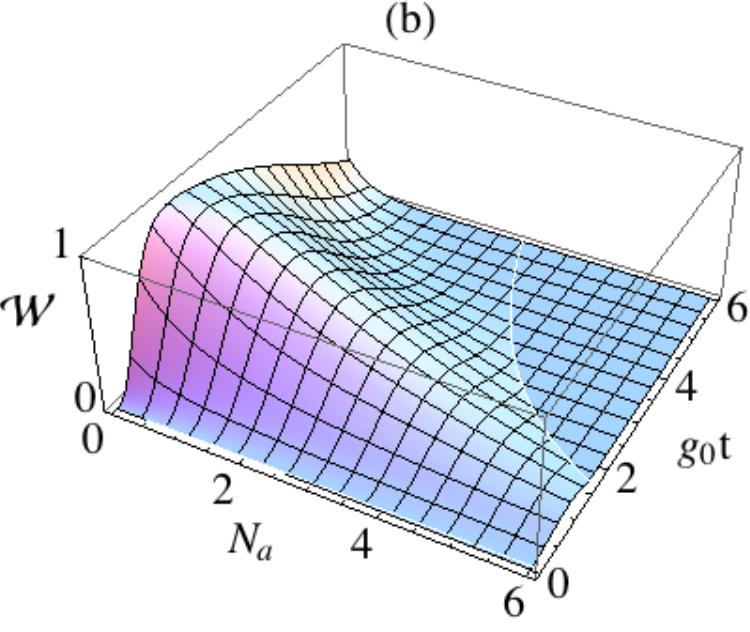}
	\end{center}
	\caption{(a): Lower bound for the negativity of the state $\rho$, quantified via the entanglement witness $\mathcal B_\mathcal{N}$. (b) Plot of $\mathcal W$, indicating the capability of the system to produce states of the form $\rho_-$, with negative Wigner functions. All quantities are plotted as functions of the oscillator average thermal excitation $N_a$ and rescaled interaction time $g_0 t$, the decoherence parameters being set to $\kappa=\gamma=0.01 g_0$. The two quantities in plots (a) and (b) show a similar qualitative behaviour, reinforcing the intuition that qubit-oscillator entanglement is a key resource for engineering non-classicality in the oscillator.\label{plots}}
\end{figure}

It Ref.~\cite{tufa-4} it was shown that the mechanism of qubit-controlled displacements is able to maximally entangle the qubit-oscillator system, even when the oscillator is initially in a highly mixed state. Our aim is to investigate how this prediction is affected by the presence of the environment. We assume that the system is initially in the separable state
\begin{equation}
	\rho_0=\kebra{+}{+}\otimes\rho_\text{th},\label{initialstate}
\end{equation}
where $\ket\pm=(\ket e\!\pm\!\ket g)/\sqrt2$ are the eigenstates of $\sigma_1$, and $\rho_\text{th}=\sum_np_n\kebra{n}{n}$ is the state of the oscillator when it is at thermal equilibrium with its environment, so that $\ket n$ are the Fock states and $p_n=N_a^n/(N_a\!+\!1)^{n+1}$. We now consider the time evolution of such a state according to the master equation \eqref{master}, and the techniques of Section~\ref{solving} can be used to provide the solution for any $t>0$. The initial conditions for the C-Matrix elements that have to be used are $\chi_j(\beta,0)=\tfrac{1}{2}{\rm e}^{-\Delta|\beta|^2},$ for $j=e,g,+,-$, and we recall that $\Delta=N_a\!+\!1/2$. After some manipulations it can be shown that the C-Matrix elements at time $t$ are given by
\begin{align}
	\chi_{eg}(\beta,t)&=\tfrac{1}{2}{\rm e}^{-\Delta|\beta+2\alpha_0|^2-w},\\
	\chi_{ge}(\beta,t)&=\tfrac{1}{2}{\rm e}^{-\Delta|\beta-2\alpha_0|^2-w},\\
	\chi_{gg}(\beta,t)&=\tfrac{1}{2}{\rm e}^{-\Delta|\beta|^2+\alpha_0\beta^*-\alpha_0^*\beta},\\
	\chi_{ee}(\beta,t)&=\tfrac{1}{2}{\rm e}^{-\Delta|\beta|^2-\alpha_0\beta^*+\alpha_0^*\beta},\\
	\alpha_0&=-i\tfrac{2g_0}{\kappa}(1-{\rm e}^{-\frac{\kappa}{2}t}),\\
	w=\gamma t&+16\Delta\left(\tfrac{g_0}{\kappa}\right)^2(\kappa t-3+4{\rm e}^{-\frac{\kappa}{2}t}-{\rm e}^{-\kappa t}),
\end{align}
where for brevity we have left implicit the functional dependence of the various parameters. These correspond to the density matrix
\begin{align}
\rho\!=\!\tfrac{1}{2}&\left[\kebra{e}{e} D(\alpha_0)\rho_\text{th} D^\dagger(\alpha_0)+\kebra{g}{g}D^\dagger(\alpha_0)\rho_\text{th} D(\alpha_0)\right]\nonumber\\
\!+\!\tfrac{{\rm e}^{-w}}{2}&\left[\kebra{e}{g}D(\alpha_0)\rho_\text{th}D(\alpha_0)+\kebra{g}{e} D^\dagger(\alpha_0)\rho_\text{th} D^\dagger(\alpha_0)\right],\label{states}
\end{align}
as it can be checked directly by substituting the above expression in Eq.~\eqref{cmatrix}. In Ref.~\cite{tufa-4}, the entanglement of the state $\rho$ was studied in the limit of large displacements ($|\alpha_0|\gg\sqrt\Delta$). Here, on the other hand, we cannot in general rely on this approximation, and we have to attack the problem without further assumptions on the parameters entering the state $\rho$. We shall quantify the qubit-oscillator entanglement via the {\it negativity} \cite{entanglo}\footnote{The negativity is defined as $\mathcal N(\rho)=||\rho^{\intercal_q}||_1-1$, where $\rho^{\intercal_q}$ is the partial transpose of $\rho$ and $||A||_1=\tr{(A^\dagger A)^{1/2}}$. An equivalent expression is $\mathcal N(\rho)=2\sum_{\lambda_-}|\lambda_-|$, where $\lambda_-$ are the negative eigenvalues of $\rho^{\intercal_q}$.}. We start by calculating the partial transpose of $\rho$ with respect to the qubit:
\begin{align}
\rho^{\intercal_q}\!=\!\tfrac{1}{2}&\left[\kebra{e}{e} D(\alpha_0)\rho_\text{th} D^\dagger(\alpha_0)+\kebra{g}{g}D^\dagger(\alpha_0)\rho_\text{th} D(\alpha_0)\right]\nonumber\\
\!+\!\tfrac{{\rm e}^{-w}}{2}&\left[\kebra{g}{e}D(\alpha_0)\rho_\text{th}D(\alpha_0)+\kebra{e}{g} D^\dagger(\alpha_0)\rho_\text{th} D^\dagger(\alpha_0)\right].\label{PT}
\end{align}
If we find a vector $\ket\psi$ such that $\bra\psi\rho^{\intercal_q}\ket\psi<0$, then the state $\rho$ must be entangled \cite{entanglo}. To this end, following \cite{tufa-4}, to each Fock state $\ket m$ we associate the test state
\begin{equation}
	\ket{\psi_m}=\tfrac{1}{\sqrt2}(\ket e D^\dagger (\alpha_0)\ket m\!-\!\ket gD(\alpha_0)\ket m).
\end{equation}
Defining $q_m=\bra{\psi_m}\rho^{\intercal_q}\ket{\psi_m}$, and using the fact that the test states are orthonormal ($\sprod{\psi_m}{\psi_{m'}}=\delta_{mm'}$), we can find a lower bound for the negativity of the form
\begin{equation}
	\mathcal N(\rho)\geq\mathcal{B}_{\mathcal{N}}(\rho)=2\sum_{q_m<0}|q_m|.\label{BN}
\end{equation}
Note that the quantity $\mathcal{B}_\mathcal{N}$ is in itself a valid entanglement witness. In particular, a positive value signals the presence of entanglement, while a value of $1$ implies that $\rho$ is a maximally entangled qubit-oscillator state. Evaluating $\mathcal B_\mathcal N(\rho)$ is straightforward once the expectation values $q_m$ entering Eq.~\eqref{BN} are known. These can be formally expressed as
\begin{align}
	q_m&=\frac{1}{2}\left[L_m\left(\!-\tfrac{\partial}{\partial N_a}\right)\tfrac{1}{N_a+1}{\rm e}^{-\tfrac{4|\alpha_0|^2}{N_a+1}}-p_m\,{\rm e}^{-w}\right],
\end{align}
where $L_m$ is the Laguerre polynomial of order $m$. The typical behaviour of $\mathcal B_N(\rho)$ is shown in Fig.~\ref{plots}(a), where we can see that the presence of qubit-oscillator entanglement can be witnessed in a range of temperatures and interaction times. 

\subsection{Preparing non-classical superposition states}
It is known that a Hamiltonian of the form \eqref{CDH} can be used to prepare non-classical superposition states of the oscillator (see e.g. \cite{solano,tufa-2,rablzoller}). Indeed, it is easy to check that by preparing a state $\ket+\otimes\ket0$, applying the time evolution operator of Eq.~\eqref{U_t}, and measuring the qubit in the basis $\ket\pm$, the oscillator is left in a superposition of coherent states of the form $\ket{\alpha_\pm}\propto\ket\alpha\pm\ket{\!-\!\alpha}$. The non-classicality of these states is typically associated to their Wigner function assuming negative values \cite{mari}. Thanks to our techniques, here we can investigate how realistic decoherence and thermalization mechanisms influence the generation of such states, and in particular the negativity of their Wigner functions. Before proceeding, we recall that given an oscillator state $\rho_a$ with characteristic function $\chi_a(\beta)=\tra{D(\beta)\rho_a}$, its Wigner function is given by \cite{q-optics}
\begin{equation}
	W_a(\alpha)=\frac{1}{\pi^2}\int{\rm d}^2\beta\,\chi_a(\beta){\rm e}^{\alpha\beta^*-\alpha^*\beta}.
\end{equation} 
Our starting point is again the separable state \eqref{initialstate}, which is the natural generalization of $\ket+\otimes\ket0$ to the case where the oscillator environment is at nonzero temperature. Following the usual strategy, we let the system evolve for a time $t$, thus obtaining the qubit-oscillator state $\rho$, as given in Eq.~\eqref{states}, and finally we measure the qubit in the basis $\ket\pm$ so that, depending on the outcome, the oscillator is projected onto the states
\begin{equation}
	\rho_\pm=(P_\pm)^{-1}\bra\pm\rho\ket\pm.\label{projected}
\end{equation}
In the above equation, $P_\pm=\tra{\bra\pm\rho\ket\pm}$ is a normalization constant, corresponding to the probability of obtaining the outcome $\ket\pm$. For simplicity we shall restrict to the outcome $\ket-$, as the Wigner function of the state $\rho_-$ exhibits the maximum negativity (if any) in zero. Correspondingly, we define the following quantity
\begin{align}
	\mathcal W&=\pi P_-\max\left\{0,-W(0)\right\},\label{W}
\end{align}
where $W(\alpha)$ is the Wigner function of $\rho_-$. Note that $\mathcal W$ encodes information about
the non-classicality of the state $\rho_-$, as well as the probability of preparing it by measuring the resource state $\rho$ [Eq.~\eqref{states}]. It may thus be interpreted as an indicator of the system's capability to produce non-classical oscillator states. The normalization constant in Eq.~\eqref{W} is chosen such that $\mathcal W$ varies between $0$ and $1$ for the considered class of states (see below). In particular, a value of $1$ signals the optimal situation in which a state with $W(0)\!=\!-2/\pi$ can be produced with probability $P_-\!=\!1/2$. Explicit calculation of the quantities involved in Eq.~\eqref{W} yields
\begin{align}
	P_-=\frac{1-{\rm e}^{-4\Delta|\alpha_0|^2-w}}{2},\\
	W(0)=\frac{{\rm e}^\frac{|\alpha_0|^2}{\Delta}-{\rm e}^{-w}}{\pi P_-}.
\end{align}
A typical plot of $\mathcal W$ is shown in Fig.~\ref{plots}(b), where we may notice a similar qualitative behavior as compared to the entanglement witness $\mathcal B_\mathcal N$ [Fig.~\ref{plots}(a)]. Indeed, in this context it is generally perceived that the possibility of generating an oscillator state with negative Wigner function is directly linked to the presence of qubit-oscillator entanglement in the state $\rho$ of Eq.~\eqref{states}. The quantitative analysis presented here seems compatible with this intuition.
\section{Conclusions}\label{fine}
We have studied a qubit-oscillator model in which a Hamiltonian generating qubit-controlled displacements competes with a thermal Markovian environment. We have shown how the model can be experimentally meaningful, in particular for nanomechanical systems, ion trap, cavity QED and circuit QED. We have shown how the dynamics of the system can be solved with minimal numerical effort, and analytically in many relevant special cases. Finally, we have applied our model and techniques to show that the system can become entangled and show non-classicality of the oscillator Wigner function even in the presence of rather strong thermal noise. Our techniques can be useful to study the experimental feasibility of many applications that rely on the mechanism of qubit-controlled displacements.
\subsection*{Acknowledgments}
We thank Sougato Bose, Myungshik Kim, Alessandro Ferraro, Gerardo Adesso and Barry Garraway for the useful discussions. We thank Mauro Paternostro for pointing out some important references. We acknowledge support from the NPRP 4-554-1-084 from Qatar National Research Fund, the UK EPSRC, the QIPIRC, the Royal Society and the Wolfson Foundation.

\end{document}